\documentclass[%
 reprint,
 amsmath,amssymb,
 aps,
  pre,
]{revtex4-2}

\usepackage{graphicx}% Include figure files
\usepackage{dcolumn}% Align table columns on decimal point
\usepackage{bm}% bold math

\usepackage{here}
\usepackage{color}
\graphicspath{{./picture/}}%pass of the file in which picture is 

\begin{document}

\preprint{APS/123-QED}

\title{Fusion of two critical points and accelerated phase dynamics in orientational ternary mixtures}% Force line breaks with \\

\author{Hiroshi Yokota}
 \email{yokotah@tmu.ac.jp}
 \altaffiliation[Also at ]{Department of Physics, Tokyo Metropolitan University.}%Lines break automatically or can be forced with \\
 
%\affiliation{%


%\affiliation{
%\affiliation{%


\date{\today}% It is always \today, today,

\begin{abstract}
Motivated by intracellular phase separation, we theoretically investigate how molecular orientation and multi-component nature affect phase behavior. 
We construct a minimal model for a ternary mixture composed of isotropic (I), anisotropic (A), and solvent (s) components by combining the Flory–Huggins and Maier–Saupe theories.
We obtain two main results from evaluating the phase behavior and the time evolution of the density fields.
First, for certain interaction parameters, two distinct binodal lines appear in the plane of the volume fractions of the I- and A-components, and merge through their respective critical points.
Second, rapid droplet formation emerges due to a weakly first-order phase transition, characterized by a discontinuity of the spinodal surface.
The first result indicates the possibility of continuous transformation between the two phase-separated states.
The second result suggests that anisotropic molecules can regulate phase separation kinetics.
These findings might be physically general beyond biological systems.
%\begin{description}
%\end{description}
\end{abstract}


\maketitle

%\tableofcontents
%
\section{Introduction}\label{sec:introduction}
Nontrivial phase behavior arising from various molecular interactions has attracted considerable attention in the field of soft matter physics.
Phase separation has also been observed even in the complicated intracellular environment composed of multiple molecular species, which is referred to as a liquid-liquid phase separation (LLPS) in cells (or intracellular phase separation).
LLPS in cells is characterized by the formation of protein-rich droplets through nucleation and growth, where the droplets are constructed by the intrinsically disordered regions (IDRs) of proteins\cite{CNT_phase_separation_in_cell1, LLPS_review}.
In biophsics, LLPS in cells is a ubiquitous phenomenon observed in many biological activities\cite{P_granule_droplet, nucleoli_droplet, FET_droplet_with_RNA_Pol, Ddx4_droplet, super_enhancer_droplet, HP1_droplet, phase_field_LLPS, CGMD_for_phase_separation, RNA_transcription_change_nuclear_body}, including pathological processes\cite{LLPS_and_disease}.
Since a protein consists of various amino acids, its phase behavior can be complicated by multiple types of molecular interactions.
It is essential to understand such a complicated phase behavior of a general phenomenon based on a unified and minimal theoretical framework.

In this study, focusing on LLPS in cells, we consider the anisotropic interactions as well as the isotropic attractions.
The LLPS in cells originates from Coulomb interaction, dipole-dipole interaction, cation-$\pi$ interaction and $\pi$-$\pi$ stacking\cite{interactions_for_phase_separation_in_cell}.
Among them, the first two are isotropic interactions, while the latter two are anisotropic.
As $\pi$ bonds specify the molecular orientation through the normal vector to the benzene ring in amino acids such as tyrosine and phenylalanine, they induce the orientational degree of freedom and the anisotropic interaction\cite{de_Gennes_Prost}.
For example, the $\pi$-$\pi$ stacking induces the nematic interaction.
In general, orientational degrees of freedom and anisotropic interactions play crucial roles in determining phase behavior influencing the shape of the phase diagram\cite{FH_MS_free_energy, MS_correct_phase_diagram}, the morphology of phase-separated structures\cite{coupling_between_density_and_orientation} and related phenomena.

Such orientational degrees of freedom --- even beyond the case of benzene rings --- are widely observed and might be indispensable for the intracellular phase separation.
For instance, cGAS and short DNA molecules ($<$ 50 nm) cooperate to form droplets \cite{DNA_induced_LL_phase_separation}.
Because short DNA ($<$ 50 nm) behaves as a rod-like molecule under physiological conditions\cite{plasmid_DNA_persistence_length}, it possesses an orientational degree of freedom associated with its molecular axis.
Moreover, orientational effects can also arise from the shape of proteins themselves, such as the cylindrical structure of histones which also cooperate with DNA to form droplets\cite{histone_droplet}. 
Furthermore, biological engineering studies point out the important role of tyrosine with a $\pi$-bond in the occurrence of intracellular phase separation\cite{role_of_tyrosine_in_LLPS}, indicating that molecular orientation is indeed a key factor in intracellular phase separation.

Intracellular phase separation essentially involves orientational effects as well as multiplicity of components.
To explain intracellular phase separation, many studies have employed the Flory-Huggins theory for the binary system composed of the droplet-forming proteins and the other molecules\cite{LLPS_review, interactions_for_phase_separation_in_cell}. 
However, in this theory, neither the multi-component nature nor the orientational effect is incorporated.
Recently, the influence of multi-component has begun to be investigated by using particle simulations\cite{MD_for_multicomponent_polymers} and constructing a theoretical model\cite{field_multi_component}.
In contrast, despite of its importance for the phase behavior, the orientational degree of freedom has not been investigated for LLPS in cells.

The aim of this study is to elucidate the phase behavior of multi-component systems with the orientational degree of freedom by constructing a unified theoretical framework.
Although, for simplicity, we present the phase behavior in an orientational ternary system in this paper,  the extension to an $N$-component system is straightforward.
We extend the conventional binary Flory-Huggins model to a ternary system composed of an isotropic protein component, an anisotropic protein component and the other molecular component.
The orientational degree of freedom is introduced by using the Maier–Saupe theory, which describes the isotropic–nematic transition within the mean field approximation.
Although our framework is motivated by intracellular phase separation, the underlying physical principles may be generally applicable to multi-component systems with the orientational degree of freedom, contributing also to engineering applications\cite{review_multi_component_LC_polymer_nano} beyond the biological context.

This paper is organized as follows.
In Sec.~\ref{sec:model}, we construct a theoretical model.
In Subsec.~\ref{sec:free_energy}, we present the free energy of a ternary system composed of an isotropic protein component, an anisotropic protein component and the other molecular component, incorporating both compositional and orientational degrees of freedom by combining Flory-Huggins theory and Maier-Saupe theory.
Based on this free energy, Subsecs.~\ref{sec:1st_order} and \ref{sec:2nd_order} describe how to evaluate the phase diagram.
In Subsec.~\ref{sec:time_evolution}, we present the time evolution equations of the density fields governed by the Cahn-Hilliard-Cook equation.
Section.~\ref{sec:result} shows the phase diagrams and their characteristics.
Moreover, by computing the dynamics of the density field, we show the coupling between the density and the orientation in our model.
In Sec.~\ref{sec:discussion}, we discuss the physical meanings of the results.
Finally, Sec.~\ref{sec:conclusion} provides concluding remarks.
\section{Model}\label{sec:model}
\subsection{Free energy}\label{sec:free_energy}
We construct a free energy of a ternary system composed of an isotropic protein component (I-component),  an anisotropic protein component (A-component) and the third component (solvent, s-component).
The free energy is described by the isotropic and anisotropic terms, which are described by the Flory-Huggins theory and the Maier-Saupe theory, respectively.

The free energy $F$ is given as
\begin{align}
  F&=N_{\rm T} f=N_{\rm T} (f^{(i)} + f^{(n)}) \label{eqn:all_free_energy}\\
  f^{(i)}&=\frac{\phi_I}{n_I}\ln{\phi_I}+\frac{\phi_A}{n_A}\ln{\phi_A}+\frac{ \phi_{\rm s} }{n_{\rm s}}\ln \phi_{\rm s} \nonumber \\
  & \ \ \ +\chi_{\rm I, A} \phi_{\rm I}\phi_{\rm A} + \chi_{\rm I, s} \phi_{\rm I} \phi_{\rm s} + \chi_{\rm A, s} \phi_{\rm A} \phi_{\rm s} \\
  f^{(n)}&=\frac{1}{2}\left( \chi_a + \frac{5}{4}\right)\phi_A^2 S^2-\frac{\phi_A}{n_A}\ln{\left(\frac{I_0\left(\Gamma_0\phi_AS\right)}{2} \right)}\\
  \Gamma_0&=\left(\chi_a + \frac{5}{4}\right)n_A\\
  I_0\left( y \right)&=\int_{-1}^1 \exp{\left[ \frac{3y}{2}  \left(x^2-\frac{1}{3} \right)\right]} dx \label{eqn:I0} \\
  S&=\frac{1}{I_0\left(\Gamma_0\phi_AS\right)} \nonumber \\
  & \ \ \ \times  \int_{-1}^1\frac{3}{2}\left(x^2-\frac{1}{3} \right) \exp{\left[ \frac{3}{2} \Gamma_0 \phi_A S \left(x^2-\frac{1}{3} \right)\right]} \label{eqn:S}.
\end{align}
Here, $f$ denotes the free energy density, and $f^{(i)}$ and $f^{(n)}$ represent its isotropic (Flory-Huggins) and anisotropic (Maier-Saupe) parts, respectively.  
$\phi_{\rm k}$ is the volume fraction of the k-component (k=$I, A, s$).
The system is assumed to be incompressible: $\phi_{\rm s} = 1-\phi_{\rm I}-\phi_{\rm A}$.
$N_{\rm k}$ is the number of chains of k-component, $n_{\rm k}$ is the number of segments per chain and $N_{\rm T}=n_{\rm I}N_{\rm I} + n_{\rm A}N_{\rm A} + n_{\rm s}N_{\rm s} $ is the total number of segments.
$\chi_{\rm kl}$ (k, l=I, A, s) denotes the Flory-Huggins interaction parameter ($\chi$-parameter) between the k-component and the l-component, and $\chi_{\rm a}$ represents the anisotropic (nematic) interaction parameter.
$S$ and $I_0$ are the orientational order parameter and the partition function for orientational degree of freedom, respectively.

Note that when the anisotropic part of the free energy density, $f^{(n)}$, is evaluated based on mean field theory, we assume that the orientational relaxation is faster than the density relaxation\cite{derivation_of_FH_MS, FH_MS_free_energy}.
Therefore, $S$ is equilibrated, and depends on the volume fraction $\phi_{\rm A}$.
$S$ is obtained self-consistently from eqns.~(\ref{eqn:I0}) and (\ref{eqn:S})\cite{MS_correct_phase_diagram}.
\subsection{Phase coexistence condition}\label{sec:1st_order}
Two types of first-order phase transitions occur: the orientational phase transition (isotropic-nematic transition, I-N transition) and the compositional phase separation.
Below, we describe how these phase boundaries are determined.
\paragraph{Isotropic-nematic transition \\ } 
The I-N transition arises from the Maier-Saupe contribution, $f^{(n)}$, in the free energy density.
The transition point is identified by the order parameter $S$ which distinguishes the isotropic state ($S=0$) from the nematic state ($S\neq 0$).
Minimization of $f^{(n)}$ with respect to $\phi_{\rm A}$ yields $S$ at the equilibrium state in the space spanned by $\phi_{\rm I}, \phi_{\rm A}$ and $\chi_{\rm k, l}.$ 
Equations.~(\ref{eqn:I0}) and (\ref{eqn:S}) are evaluated numerically using Simpson's rule.
\paragraph{compositional phase separation via nucleation and growth \\ }
The coexistence condition between k-rich and k-poor phases (the binodal surface) is derived by equating the chemical potentials of all components.
The binodal surface is plotted in the $(\phi_{\rm I}, \phi_{\rm A},  \chi_{\rm k, l})$ space.

The chemical potentials, $\mu_{\rm k}=\partial F/(\partial N_{\rm k})$, are expressed as
\begin{align}
  \mu_{\rm I}\left(\phi_{\rm I}, \phi_{\rm A}  \right)&=\ln \phi_{\rm I} + \left(1-\frac{n_{\rm I}}{n_{\rm A}} \right)\phi_{\rm A}+\left(1-\frac{n_{\rm I}}{n_{\rm s}} \right)\phi_{\rm s} \nonumber \\
  & \ \ \ + \chi_{\rm I, A}n_{\rm I}\phi_{\rm A}^2 + \chi_{\rm I, s}n_{\rm I}\phi_{\rm s}^2  \nonumber \\
  & \ \ \ + (\chi_{\rm I, A} + \chi_{\rm I, s}- \chi_{\rm A, s})n_{\rm I}\phi_{\rm A}\phi_{\rm s} \nonumber \\
  & \ \ \ + \frac{1}{2} \left(\chi_{\rm a} + \frac{5}{4} \right)n_{\rm I}\phi_{\rm A}^2 S^2\\
  \mu_{\rm A}\left(\phi_{\rm I}, \phi_{\rm A}  \right)&=\ln \phi_{\rm A}+\left(1-\frac{n_{\rm A}}{n_{\rm I}} \right)\phi_{\rm I}+\left(1-\frac{n_{\rm A}}{n_{\rm s}} \right)\phi_{\rm s} \nonumber \\
  & \ \ \ +\chi_{\rm I, A}n_{\rm A}\phi_{\rm I}^2 + \chi_{\rm A, s}n_{\rm A}\phi_{\rm s}^2 \nonumber \\
  & \ \ \ +(\chi_{\rm I, A} + \chi_{\rm A, s} - \chi_{\rm I, s})n_{\rm A}\phi_{\rm I}\phi_{\rm s}\nonumber \\
  & \ \ \ + \frac{1}{2}\left(\chi_{\rm a} + \frac{5}{4} \right)n_{\rm A}\phi_{\rm A}^2S^2 -\ln \left[\frac{I_0(\Gamma_0 \phi_{\rm A}S)}{2} \right]\\
 \mu_{\rm s}\left(\phi_{\rm I}, \phi_{\rm A}  \right)&=\ln \phi_{\rm s} + \left(1-\frac{n_{\rm s}}{n_{\rm I}} \right)\phi_{\rm I}+\left(1-\frac{n_{\rm s}}{n_{\rm A}} \right)\phi_{\rm A} \nonumber \\
  & \ \ \ + \chi_{\rm I, s}n_{\rm s}\phi_{\rm I}^2 + \chi_{\rm A, s}n_{\rm s}\phi_{\rm A}^2 \nonumber \\
  & \ \ \  + (\chi_{\rm I, s} + \chi_{\rm A, s}- \chi_{\rm I, A})n_{\rm s}\phi_{\rm I}\phi_{\rm A} \nonumber \\
  & \ \ \ + \frac{1}{2} \left(\chi_{\rm a} + \frac{5}{4} \right)n_{\rm s}\phi_{\rm A}^2 S^2.
\end{align}
At coexistence between phases 1 and 2, the following conditions are satisfied:
\begin{align}
  \mu_{\rm I}\left(\phi_{\rm I}^{(1)}, \phi_{\rm A}^{(1)}\right)&=\mu_{\rm I}\left(\phi_{\rm I}^{(2)}, \phi_{\rm A}^{(2)} \right) \label{eqn:binodal_I}\\
  %
  %
  %4
  %
  \mu_{\rm A}\left(\phi_{\rm I}^{(1)}, \phi_{\rm A}^{(1)} \right)&=\mu_{\rm A}\left(\phi_{\rm I}^{(2)}, \phi_{\rm A}^{(2)} \right)\label{eqn:binodal_A}\\
  \mu_{\rm s}\left(\phi_{\rm I}^{(1)}, \phi_{\rm A}^{(1)} \right)&=\mu_{\rm s}\left(\phi_{\rm I}^{(2)}, \phi_{\rm A}^{(2)} \right).\label{eqn:binodal_s}
\end{align}
Here, $\phi_{\rm I}^{(1)}$ specifies the volume fraction of the I-component in the phase 1.
The other variables are also defined in the same manner.

To determine the binodal surface in the $(\phi_{\rm I}, \phi_{\rm A}, \chi_{\rm k, l})$ space (the binodal line on the $\phi_{\rm I}$-$\phi_{\rm A}$ plane for fixed $\chi_{\rm k, l}$), we define a loss function $L (\phi_{\rm I}^{(1)}, \phi_{\rm A}^{(1)}, \phi_{\rm I}^{(2)}, \phi_{\rm A}^{(2)})$.
\begin{align}
  L(\phi_{\rm I}^{(1)}, \phi_{\rm A}^{(1)}, \phi_{\rm I}^{(2)}, \phi_{\rm A}^{(2)})&=\frac{1}{2}\left(f^2 + g^2+h^2 \right),
\end{align}
where
\begin{align}
  f(\phi_{\rm I}^{(1)}, \phi_{\rm A}^{(1)}, \phi_{\rm I}^{(2)}, \phi_{\rm A}^{(2)})&=\mu_{\rm I}\left(\phi_{\rm I}^{(1)}, \phi_{\rm A}^{(1)} \right)-\mu_{\rm I}\left(\phi_{\rm I}^{(2)}, \phi_{\rm A}^{(2)} \right)\\
  g(\phi_{\rm I}^{(1)}, \phi_{\rm A}^{(1)}, \phi_{\rm I}^{(2)}, \phi_{\rm A}^{(2)})&=\mu_{\rm A}\left(\phi_{\rm I}^{(1)}, \phi_{\rm A}^{(1)} \right)-\mu_{\rm A}\left(\phi_{\rm I}^{(2)}, \phi_{\rm A}^{(2)}\right)\\
  h(\phi_{\rm I}^{(1)}, \phi_{\rm A}^{(1)}, \phi_{\rm I}^{(2)}, \phi_{\rm A}^{(2)})&=\mu_{\rm s}\left(\phi_{\rm I}^{(1)}, \phi_{\rm A}^{(1)} \right)-\mu_{\rm s}\left(\phi_{\rm I}^{(2)}, \phi_{\rm A}^{(2)} \right)
\end{align}
Since the binodal surface is characterized by $L=0$, we evaluate the binodal surface (or line) by minimizing $L$ with respect to $\phi_{\rm I}^{(1)}, \phi_{\rm A}^{(1)}, \phi_{\rm I}^{(2)}$ and $\phi_{\rm A}^{(2)}$ using the Newton method\cite{minimazation_method}.
\subsection{Condition for thermal instability}\label{sec:2nd_order}
The spinodal decomposition is classified into a second order phase transition resulting from the thermal instability of a certain state (for example, a uniform state).
In this system it is characterized by the spinodal surface in the $(\phi_{\rm I}, \phi_{\rm A}, \chi_{\rm k, l})$ space.

The spinodal surface is determined by the smallest eigenvalue of Hessian matrix of $f(\phi_{\rm I}, \phi_{\rm A})$ achieving zero.
Here, the Hessian matrix $H$ is evaluated as
\begin{align}
  H&=\left(
  \begin{array}{cc}
    \frac{\displaystyle \partial^2 f}{\displaystyle \partial \phi_{\rm I}^2} & \frac{\displaystyle \partial^2 f}{\displaystyle \partial \phi_{\rm A} \partial \phi_{\rm I}}\\
    \frac{\displaystyle \partial^2 f}{\displaystyle \partial \phi_{\rm A} \partial \phi_{\rm I}} &\frac{\displaystyle \partial^2 f}{\displaystyle \partial \phi_{\rm A}^2}
  \end{array}
  \right). \label{eqn:Hessian}
\end{align}
The components are given by 
\begin{align}
  \frac{\displaystyle \partial^2 f}{\displaystyle \partial \phi_{\rm I}^2}&=\frac{1}{n_{\rm I}\phi_{\rm I}} +\frac{1}{n_{\rm s}(1-\phi_{\rm s}-\phi_{\rm A})} -2 \chi \\
  \frac{\displaystyle \partial^2 f}{\displaystyle \partial \phi_{\rm A} \partial \phi_{\rm I}}&=\frac{1}{n_{\rm s} (1-\phi_{\rm I}-\phi_{\rm A})}-\chi\\
  \frac{\displaystyle \partial^2 f}{\displaystyle \partial \phi_{\rm A}^2}&=\frac{1}{n_{\rm A}\phi_{\rm A}} +\frac{1}{n_{\rm s}(1-\phi_{\rm I}-\phi_{\rm A})} -2 \chi \nonumber \\
  &-\frac{\Gamma_0 S^2 }{n_{\rm A}} \frac{\displaystyle 1 }{\displaystyle \left\{1 - \Gamma_0 \phi_{\rm A}\left(\langle P_2(x)^2 \rangle_S- \langle P_2(x) \rangle_S^2 \right) \right\}} ,\label{eqn:second_derivative_of_f_phi_A}
\end{align}
where $P_2(x)=3(x^2-1/3)/2$ and $\langle \cdots \rangle_S$ means the average of a physical quantity $\cdots$ for the orientation distribution.
For instance,  $S=\langle P_2(x) \rangle_S$.
The spinodal surface is obtained by finding $\chi_{\rm kl, s}$ satisfying ${\rm det}H = 0$, using the Newton-Raphson method.
\subsection{Time evolution of density field}\label{sec:time_evolution}
Based on the free energy functional extended from eqn.~(\ref{eqn:all_free_energy}), we describe the time evolution of the density fields $\phi_{\rm I}$ and $\phi_{\rm A}$ by using the Cahn-Hilliard-Cook equation.
The free energy functional $F$ is
\begin{align}
  F[\phi_{\rm I}(\mbox{\boldmath$r$}), \phi_{\rm A}(\mbox{\boldmath$r$})]&=\int d\mbox{\boldmath$r$} \left[f (\phi_{\rm I}(\mbox{\boldmath$r$}), \phi_{\rm A}(\mbox{\boldmath$r$}))
  +\frac{\kappa_{\rm I}}{2}|\nabla \phi_{\rm I}|^2 + \frac{\kappa_{\rm A}}{2}|\nabla \phi_{\rm A}|^2\right],
\end{align}
where $\kappa_{\rm k}$  (k=I or A) denotes the gradient energy coefficient.
The time evolution of $\phi_{\rm k}$ is governed by  
\begin{align}
  \frac{\partial \phi_{\rm k}}{\partial t}(\mbox{\boldmath$r$}, t)&=M\nabla^2 \frac{\delta F}{\delta \phi_{\rm k}} + \zeta(\mbox{\boldmath$r$}, t), \label{eqn:CHC}
\end{align}
where $M$ is the mobility, $\zeta(\mbox{\boldmath$r$}, t)$ is a Gaussian white noise representing thermal fluctuations.
$\zeta (\mbox{\boldmath$r$}, t)$ satisfies that 
\begin{align}
  \langle \zeta (\mbox{\boldmath$r$}, t) \zeta (\mbox{\boldmath$r$}^\prime, t^\prime) \rangle&=-M k_{\rm B}T\nabla^2 \delta(\mbox{\boldmath$r$}-\mbox{\boldmath$r$}^\prime)\delta(t-t^\prime), \label{eqn:power_spectrum_thermal_fluctuation}
\end{align}
where $\langle \cdots \rangle$ means the ensemble average of a physical quantity $\cdots$.
We numerically solve eqn.~(\ref{eqn:CHC}) under periodic boundary conditions using the semi-implicit Fourier-spectral method\cite{semi_implicit_TDGL}.

\section{Results}\label{sec:result}
\subsection{parameters}
In this paper, for simplicity, we mainly use a representative parameter set in which most parameters are fixed, while the interaction parameter $\chi \equiv \chi_{\rm IA}=\chi_{\rm Is}=\chi_{\rm As}$ is a control parameter.
Here, we assume $\chi_{\rm IA}=\chi_{\rm Is}=\chi_{\rm As}$, since asymmetric interaction parameters do not qualitatively affect the essential nature of our system.
Based on the Flory-Huggins-Maier-Saupe theory for a binary system\cite{derivation_of_FH_MS}, the nematic interaction parameter is given by 
\begin{align}
  \chi_{\rm a} &= \frac{5\chi}{2}. \label{eqn:chia}
\end{align}
We also use $n_{\rm I}=25$, $ n_{\rm A}=2$ and $n_{\rm s}=1$.

For the time evolution of the density field, the simulations on a two-dimensional space are performed by using the following parameters:
system size $N=512$ (per an axis),  spatial mesh size $\Delta x = \Delta y = 0.10$, time step size $\Delta t = 1.0\times 10^{-3}$, mobility $M=2.0$, and gradient energy coefficients $\kappa=\kappa_{\rm I} = \kappa_{\rm A} = 0.10$. 
The thermal energy is set to $k_{\rm B}T=0.10/\chi$.

\subsection{Criterion for phase separation via first-order phase transition}
Our system exhibits both the I-N transition and the compositional phase separation between k-rich and k-poor phases (k, l=I, A, s) depending on $\chi$.
\paragraph{Isotropic-Nematic transition \\ } 
\begin{figure}[tb]
  \begin{center}
    \includegraphics[width=7cm]{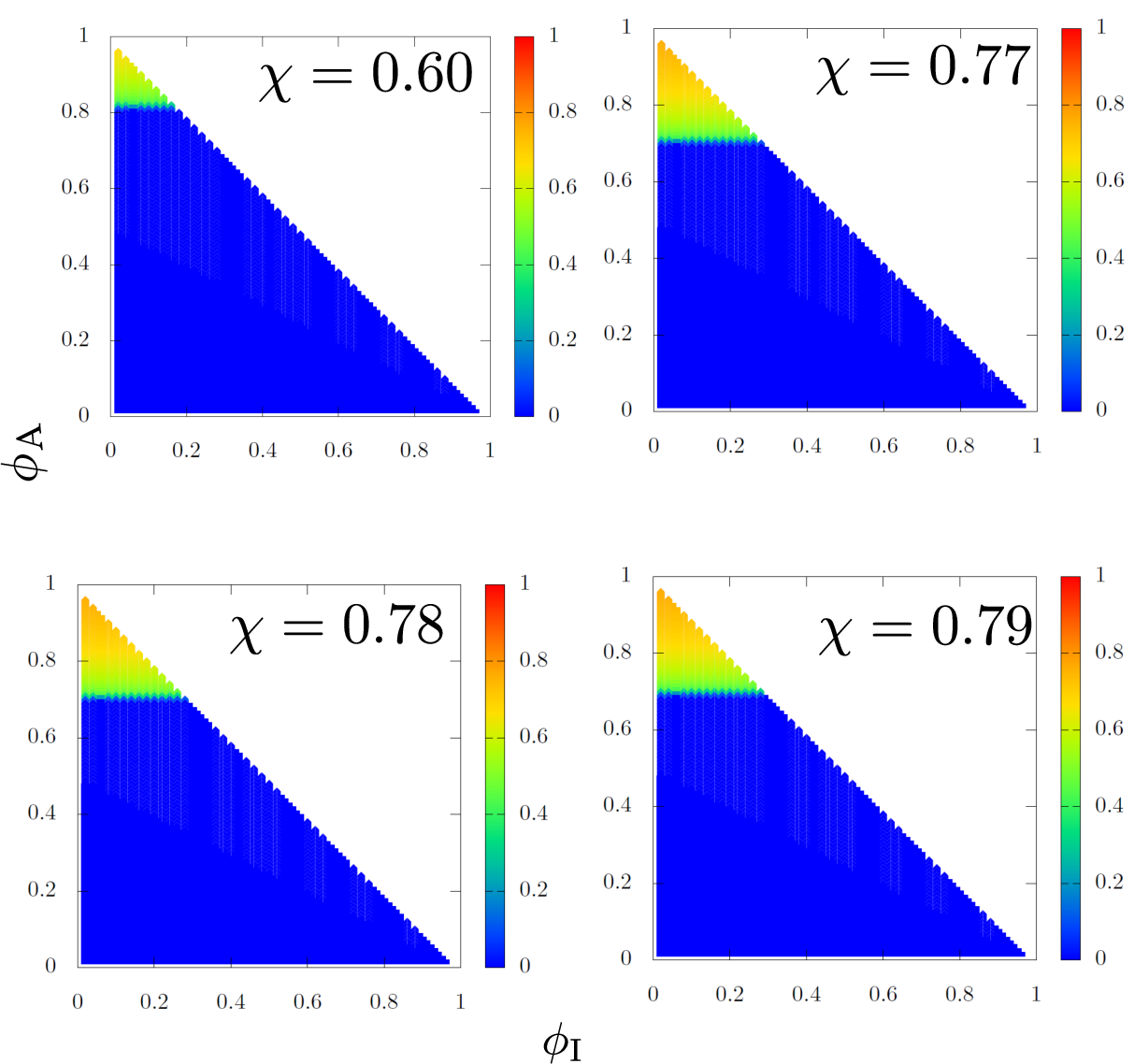}
    \caption{Color plot of the orientation order parameter $S$ for $\chi \in[0.60:0.79]$.
    The horizontal and vertical axes indicate $\phi_{\rm I}$ and $\phi_{\rm A}$, respectively.
    }
    \label{fig:orientation_large_chi}
 \end{center}
\end{figure}

We evaluate the orientation order parameter $S$ on the $\phi_{\rm I}$-$\phi_{\rm A}$ plane for various values of $\chi$ as shown in Fig.~\ref{fig:orientation_large_chi}.
The I-N transition point $\chi_a^{\rm (I-N)}$ is identified as the value of $\chi$ where $S$ first becomes nonzero (i.e., where nematic ordering appears).

The nematic phase first appears at $\chi = 0.42$, although the corresponding plot is not shown because the change is too subtle to be visible in the $\phi_{\rm I}$–$\phi_{\rm A}$ plane within the range $0 \le \phi_{\rm I}, \phi_{\rm A} \le 1$.
The corresponding transition point is defined as
\begin{align}
  \chi_a^{\rm (I-N)} + \frac{5}{4}&=\frac{5}{2} \times 0.42 + \frac{5}{4}  =2.30.
\end{align}

For $\chi \ge 0.42$ $(\chi_a\ge \chi_a^{\rm (I-N)})$, the nematic region in the $\phi_{\rm I}$-$\phi_{\rm A}$ plane expands as $\chi$ increases.
\paragraph{Compositional phase separation between k-rich and k-poor phases (k = I, A, s) \\ } 
\begin{figure}[tb]
  \begin{center}
    \includegraphics[width=7cm]{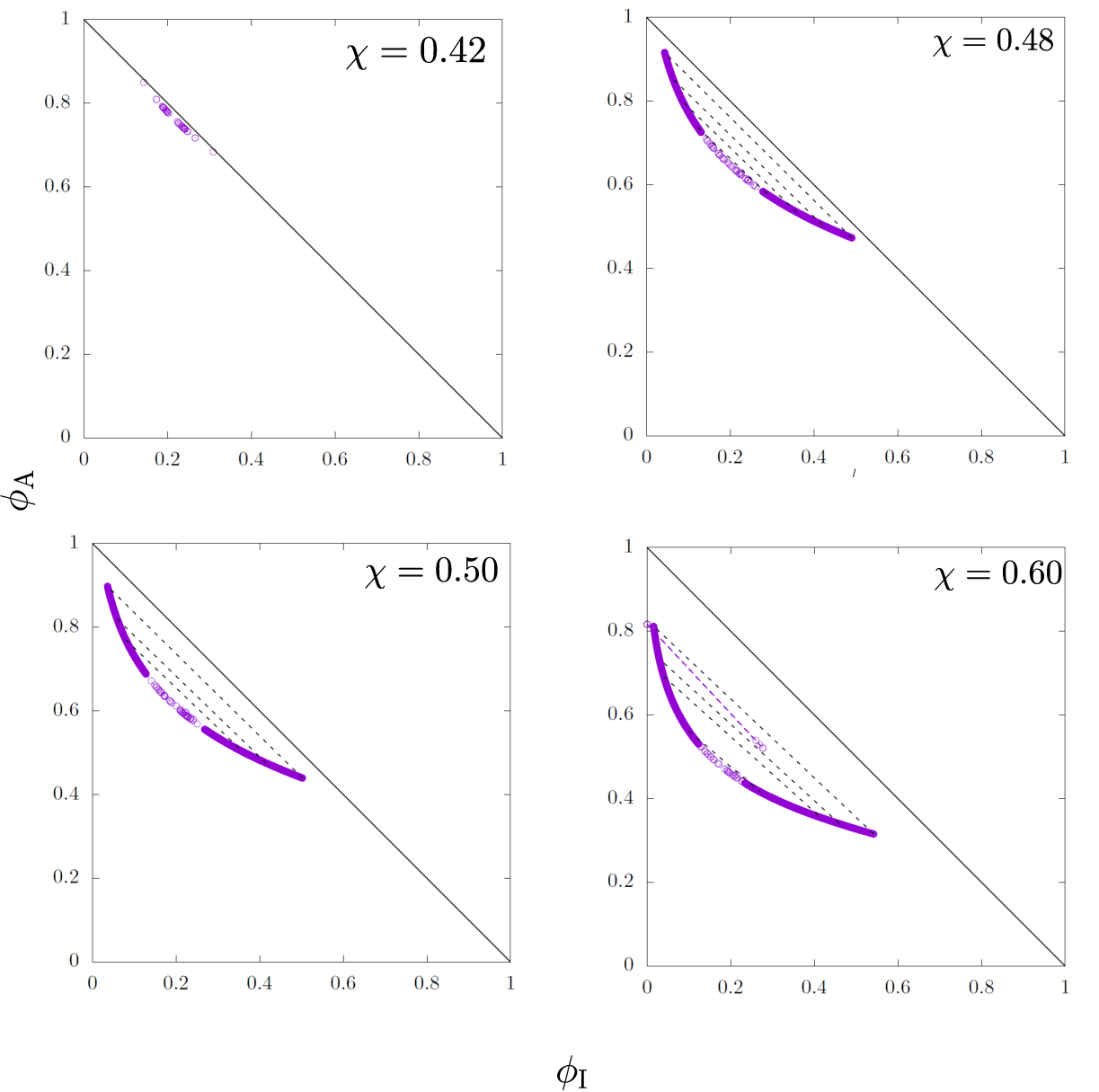}
    \caption{Binodal lines on the $\phi_{\rm I}$ -$\phi_{\rm A}$ plane for $\chi \in[0.42:0.60]$.
    The horizontal and vertical axes indicate $\phi_{\rm I}$ and $\phi_{\rm A}$, respectively.
    Purple symbols indicate the binodal points; the black dashed line indicates the tie line for the I-I phase separation, and the purple dashed line shows the tie line for isotropic–nematic coexistence.
    The black solid line corresponds to $\phi_{\rm I} + \phi_{\rm A} = 1$.
    }
    \label{fig:phase_diagram_small_chi}
 \end{center}
\end{figure}
\begin{figure}[tb]
  \begin{center}
    \includegraphics[width=7cm]{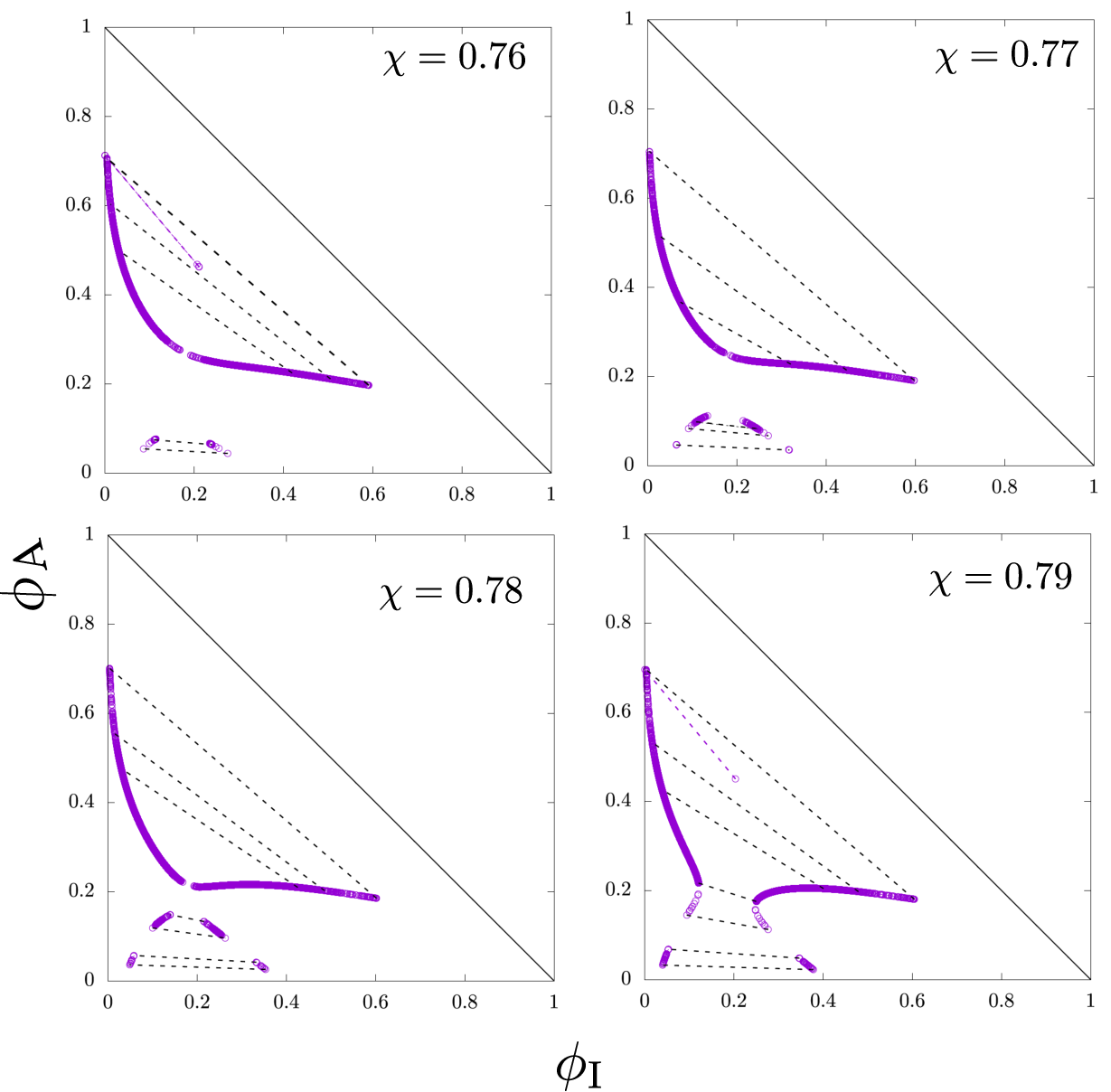}
    \caption{Binodal lines for $\chi \in [0.76, 0.79]$.
    Axes and symbols are the same as in Fig.~\ref{fig:phase_diagram_small_chi}.
    For $\chi > 0.76$, two binodal lines appear: an upper one (I-rich and A-rich) and a lower one (I-rich and s-rich).}
    \label{fig:phase_diagram_large_chi}
 \end{center}
\end{figure}
Figures~\ref{fig:phase_diagram_small_chi} and \ref{fig:phase_diagram_large_chi} show the binodal lines on the $\phi_{\rm I}$-$\phi_{\rm A}$ plane at fixed values of $\chi$.
The first isotropic-isotropic (I-I) phase separation appears at $\chi=0.42$.
Because this binodal line is close to $\phi_{\rm I}+\phi_{\rm A} = 1$, it represents the phase separation between the isotropic I-rich and isotropic A-rich phases.
This is more clearly seen in the triangular phase diagram in Appendix~\ref{sec:triangular_phase_diagram}.
The coexistence region grows with the increase in $\chi$.

For $\chi > 0.50$, an additional binodal line appears, corresponding to the coexistence between isotropic and nematic phases (purple dashed line).
One end of its tie line has composition $\phi_{\rm I}<0.010$ and $\phi_{\rm A}>0.80$, corresponding to an almost binary mixture composed of the A-component and the s-component.
Since this composition lies within the nematic region (see Fig.~\ref{fig:orientation_large_chi}), this binodal line represents the coexistence between the isotropic and nematic phases.
This is consistent with the known phase behavior in polymer–liquid-crystal mixtures\cite{derivation_of_FH_MS}.
Note that the isotropic-nematic coexistence line is not detected for $\chi=0.77, 0.78$ which is due to the limited numerical precision.
Since this issue is not central to the main results of the present study, we do not elaborate on it further.

At $\chi = 0.76$, a new phase separation emerges between the I-rich and s-rich phases (see Fig.~\ref{fig:phase_diagram_large_chi}), corresponding to that predicted by the conventional binary Flory-Huggins model for LLPS in cells.
Thus, the inclusion of the anisotropic component broadens the overall phase-separated region compared with that in previous binary models.

As $\chi$ increases, the A-component fraction in the I–s binodal line increases, while that in the I–A binodal line decreases.
At approximately  $(\phi_{\rm I},  \phi_{\rm A}, \chi)\simeq (0.20,  0.20, 0.79)$, the two binodal lines merge at their vertices corresponding to the critical points in these phase-separated regions.
We refer to this merged point as the “merged critical point” in this paper.
\subsection{Criterion for phase separation via second order phase transition}
\begin{figure}[tb]
  \begin{center}
    \includegraphics[width=8cm]{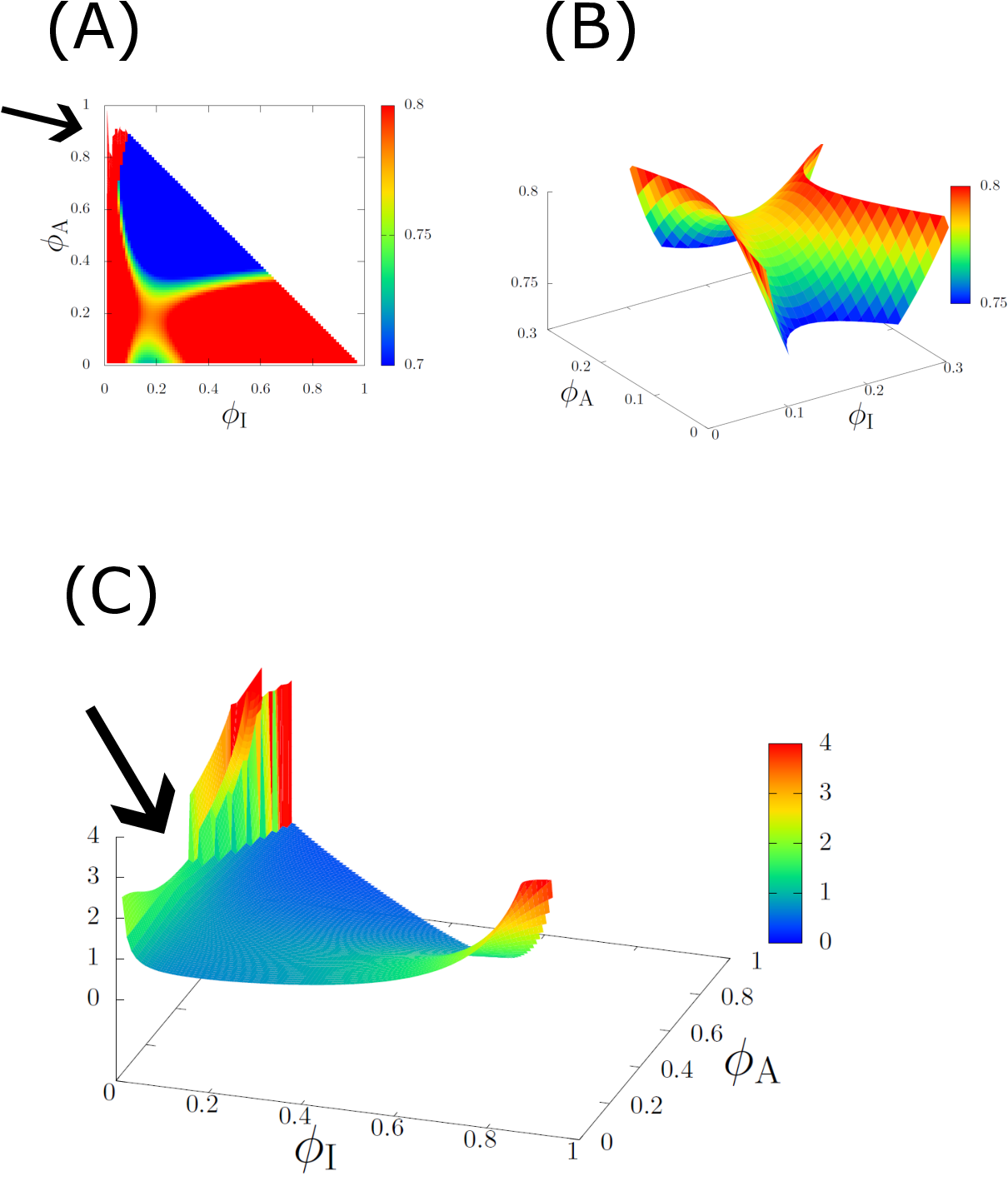}
    \caption{Spinodal surface.
    (A) Color map of spinodal surface $\chi_{\rm s}$ on the $\phi_{\rm I}$-$\phi_{\rm A}$ plane.
    The red color specifies high $\chi_{\rm s}$, while blue color means its low value.
    (B) 3D view around  the merged critical point.
    (C) Full 3D view of the spinodal surface. 
    Black arrows indicate discontinuity in $\chi_{\rm s}$.
    }
    \label{fig:spinodal}
 \end{center}
\end{figure}
The spinodal surface $\chi_{\rm s} (\phi_{\rm I}, \phi_{\rm s})$ is shown in Fig.~\ref{fig:spinodal}.
It exhibits the saddle point around $(\phi_{\rm I}, \phi_{\rm A}, \chi_{\rm s})=(0.20, 0.20, 0.79)$ (Fig.~\ref{fig:spinodal} (A), (B)), which corresponds to the merged critical point identified in Fig.~\ref{fig:phase_diagram_large_chi}.
Thus, the saddle point on the spinodal surface characterizes the merged critical point.

In the region with large $\phi_{\rm A}$ and small $\phi_{\rm I}$, $\chi_{\rm s}$ shows the discontinuous behavior (Fig.~\ref{fig:spinodal}(A),(C)).
This discontinuity arises from the I–N transition, where the discontinuous change of $S$ induces the discontinuities in $\partial_{\phi_{\rm A}}^2 f$ (see Eq.~(\ref{eqn:second_derivative_of_f_phi_A})).
\subsection{Time evolution of the density field of A-component}
We numerically solve the Cahn-Hilliard-Cook equation (\ref{eqn:CHC}) with $\chi=0.75$ starting with the initial values $\phi_{\rm I}=0.10 \pm u, \phi_{\rm A}=0.70 \pm u$, where $u$ is a uniform random variable distributed on $[-0.025, 0.025]$.
This initial condition corresponds to a point near the discontinuous region of the spinodal surface.

\begin{figure}[tb]
  \begin{center}
    \includegraphics[width=7cm]{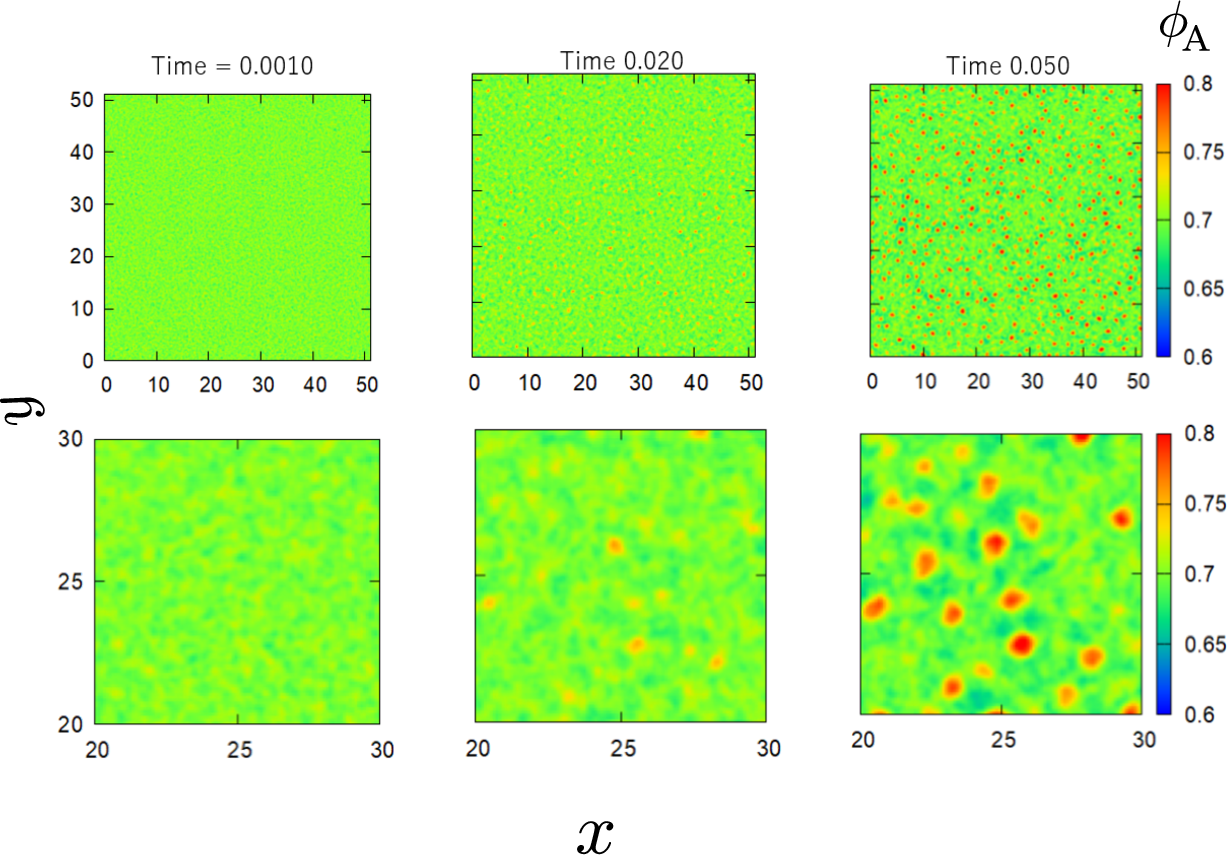}
    \caption{Time evolution of the A-component density field $\phi_{\rm A}$.
    The upper panels show the full system; the lower panels show magnified views highlighting droplet formation.
    }
    \label{fig:droplet_formation}
 \end{center}
\end{figure}

The time evolution of the A-component density field is shown in Fig.~\ref{fig:droplet_formation}.
Droplets form rapidly at $t = 0.050$, much faster than typical I-I spinodal decomposition.
For comparison, Appendix~\ref{sec:typical_spinodal} shows the time evolution for a ternary Flory–Huggins system, where density fluctuations appear only after $t = 5.0$.
Thus, the droplet formation observed here is roughly 100 times faster.

This rapid droplet formation requires the thermal fluctuation and thus corresponds to a first-order  phase transition.
Note that the same simulation without the stochastic term ({\it i.e.}, using the deterministic Cahn–Hilliard equation) does not exhibit such droplet formation.

\begin{figure}[tb]
  \begin{center}
    \includegraphics[width=9cm]{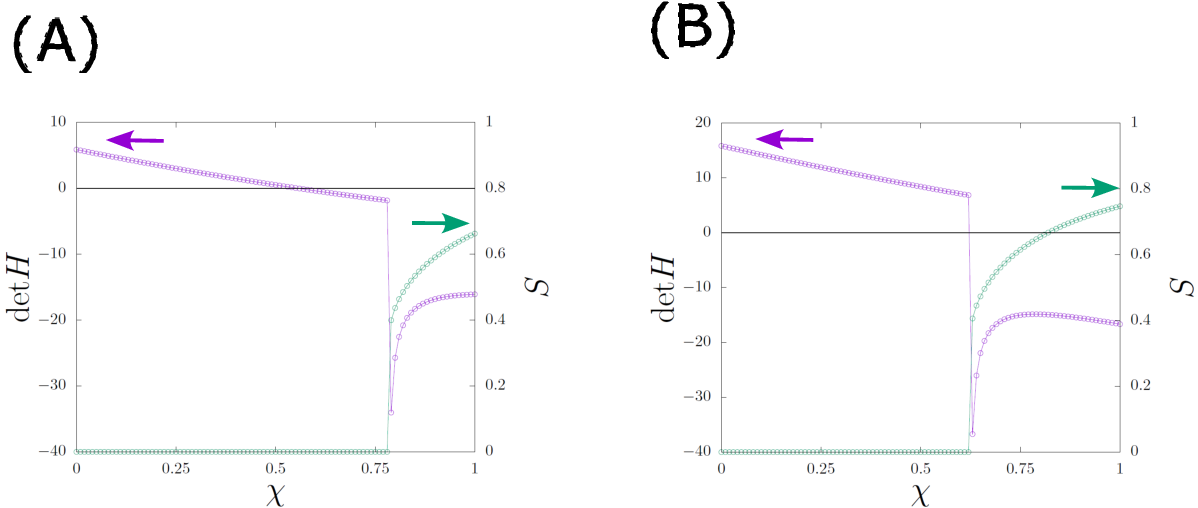}
    \caption{$\chi$-dependence of $\det H$ (purple) and $S$ (green).
    Left axis: $\det H$; right axis: $S$.
    The black horizontal line denotes $\det H = 0$.
    (A) $(\phi_{\rm I}, \phi_{\rm A}) = (0.10, 0.70)$; (B) $(\phi_{\rm I}, \phi_{\rm A}) = (0.020, 0.80)$.
    }
    \label{fig:detH_and_S_manyA}
 \end{center}
\end{figure}

The rapid droplet formation originates from the coupling between the density and the orientation.
To determine this, we compute det$H$ and $S$ as functions of $\chi$ at $(\phi_{\rm I}, \phi_{\rm A})=(0.10, 0.70)$ (Fig.~\ref{fig:detH_and_S_manyA} (A)).
det$H$ becomes 0 at $\chi \simeq 0.55$, indicating the large amplitude of density fluctuation.
This large density fluctuation leads to a spinodal decomposition.
Around $\chi = 0.75$, $\det H$ abruptly decreases to a large negative value ($\det H \sim -35$), while $S$ simultaneously jumps from 0 to a finite value.
This behavior indicates a large-amplitude density fluctuation driven by the density-orientation coupling, which effectively achieves the dynamic characteristics of a deep quench.
We therefore term it a “pseudo deep quench”.

When $\phi_{\rm A}$ reaches the pseudo deep quench region due to the thermal fluctuation, the large amplitude of density fluctuation and the I-N transition cooperate to drive the system from a miscible state to a phase separated state in the droplet formation manner through the first-order phase transition.
Note that, during the rapid droplet formation, a certain Fourier mode of density fluctuations increases with time, as in the early stage of the spinodal decomposition (see Appendix~\ref{sec:power_spectra}).
\section{Discussion}\label{sec:discussion}
In this paper, we constructed a theoretical model combining the Flory-Huggins theory and the Maier-Saupe theory in a ternary system composed of the isotropic protein component (I-component), the anisotropic protein component (A-component) and the other molecular component (solvent, s-component).
In our model, we found two binodal lines: one between the I-rich and A-rich phases, and the other between the I-rich and s-rich phases.
These two binodal lines approach each other as $\chi$ increases, and when $\chi$ reaches a certain value, they merge through their respective critical points.
The ``merged critical point'' corresponds to the saddle point on the spinodal surface. 
Moreover, in the region containing a large amount of A-component, the spinodal surface exhibits the discontinuous behavior.
Around this region, the droplet formation proceeds more rapidly than the typical spinodal decomposition predicted by the Flory-Huggins theory.
This accelerated dynamics arises from the cooperation between the first-order phase transition and the density fluctuation.

The present results can be interpreted as arising from two main contributions:
\begin{enumerate}
  \item the effect of the ternary nature.
  \item the effect of the orientational degree of freedom.
\end{enumerate}
The ``merged critical point'', corresponding to the saddle point of the spinodal surface, originates from the ternary nature of the system.
While the previous theoretical work\cite{three_phase_coexistence_binodal_fusion} demonstrates that the fusion of a critical point and another binodal line leads to three-phase coexistence, our merged critical point characterizes the fusion between two critical points, which instead extends the two-phase coexistence region.
Once the merged critical point appears, continuous pathways exist between the I–A and I–s phase-separated regions, along which thermodynamic quantities vary smoothly without divergence.
Such continuous pathways among phase-separated regions may also appear in multi-component systems, since these systems generally possess multiple binodal lines (surfaces).

On the other hand, the ``pseudo deep quench''  (associated with the rapid droplet formation) arises from the effect of the orientational degree of freedom.
This phenomenon directly causes the discontinuity of the spinodal surface, as shown in Fig.~\ref{fig:spinodal}(C).
Figure~\ref{fig:detH_and_S_manyA}(B) shows an abrupt drop in ${\rm det}H$ from a positive region to a negative region.
Since the spinodal condition is characterized by ${\rm det}H = 0$, such an abrupt change may lead to the false detection of the actual spinodal point.
Because the pseudo deep quench characterizes the first-order phase transition involving the large density fluctuation derived from the I–N transition, it can be interpreted as a weakly first-order phase transition\cite{pretransition} that drives rapid droplet formation.
Because this weakly-first-order phase transition is intrinsic to the Flory-Huggins-Maier-Saupe theory, the rapid droplet formation predicted here may represent a general phenomenon, not limited to biological systems.

The orientational effect suggests the possibility to control the rate of phase separation.
For instance, in the context of LLPS in cells, the overexpression of anisotropic molecules accelerates droplet formation.
Examples of such anisotropic molecules include tyrosine, short DNA (shorter than 50 nm), and cylindrical proteins such as histones.
Note that the rapid droplet formation coming from a weakly first-order phase transition is physically general and applicable to multi-component systems with orientational degrees of freedom, beyond biological context.

Furthermore, the ternary nature leads to continuous pathways connecting two phase-separated regions beyond LLPS in cells.
A ternary system with two phase-separated regions (I–A and I–s) can be continuously transformed from one phase-separated region to another when the two binodal lines merge at a certain $\chi$.
Along such a pathway, some physical quantities --- such as thermal conductivity, internal energy, and free energy --- can be described as continuous functions\cite{CL}.
Because this pathway resembles the supercritical region that exhibits a dynamical crossover\cite{Frenkel_line_supercritical_from_heat_capacity}, a similar crossover behavior may occur in multi-component systems as well.
\section{Conclusion}\label{sec:conclusion}
Motivated by intracellular phase separation and its underlying interactions, we constructed a theoretical model to explore the phase behavior in a ternary system with an orientational degree of freedom.
Our model combines the Flory-Huggins theory for a ternary mixture with the Maier-Sauper theory to incorporate  the orientational interaction.
In this framework, both isotropic and anisotropic interactions are characterized by the interaction parameter $\chi$.
Although inspired by intracellular phase separation, the resulting phenomena are physically general and extend beyond biological systems.

Using this model, we evaluated the binodal surface, the spinodal surface, and the time evolution of the density field.
We found the two phase-separated regions between isotropic phases and continuous pathways connecting them at a certain $\chi$ value.
Such pathways can also exist in multi-component systems beyond ternary mixtures.

We also found that the rapid droplet formation in our model originates from the weakly first-order phase transition.
The “pseudo deep quench” characterizes this rapid formation, triggered by large density fluctuations driven by the I–N transition.
This behavior reflects a general property of weakly first-order transitions.

These results suggest possible strategies for (i) investigating the behavior of physical quantities across multiple phase-separated regions in multi-component systems and (ii) controlling the kinetics of phase separation with orientational degrees of freedom.
The first aspect provides a thermodynamic framework for understanding physical quantities across multiple phases, while the second implies that anisotropic molecules --- such as short DNA, cylindrical protein complexes, or aromatic amino acids --- could act as kinetic modulators of droplet formation.

Finally, since our present model assumes that the relaxation of $S$ is faster than that of $\phi_{\rm k}$ (k= I, A, s), the dynamical coupling between the density and the orientation could not be addressed.
In future work, we will focus on the coupled relaxation dynamics of the density and the orientation at the early stage of phase behavior.
\begin{acknowledgments}
The author is indebted to Professor T. Sakaue and Mr. N. Iso for valuable discussions.
This work is supported by the Grant-in-Aid for Scientific Research (Grant Number 22K14018) from The Ministry of Education, Culture, Sports, Science and Technology(MEXT), Japan.
\end{acknowledgments}
\appendix

\section{Triangular phase diagram corresponding to Figs.~\ref{fig:phase_diagram_small_chi}, \ref{fig:phase_diagram_large_chi} and \ref{fig:spinodal}}\label{sec:triangular_phase_diagram}
\begin{figure}[htb]
  \begin{center}
    \includegraphics[width=7cm]{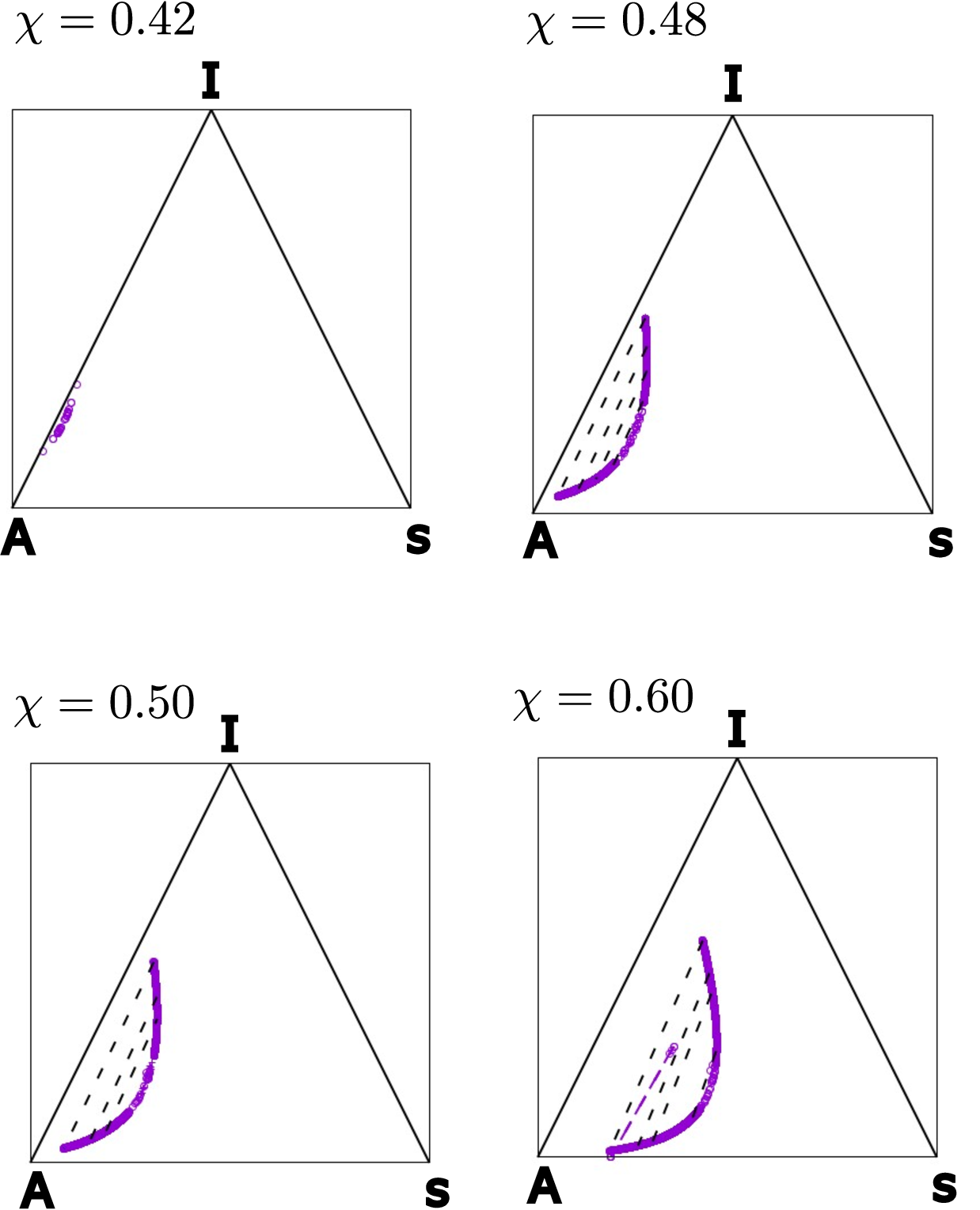}
    \caption{Binodal line described on triangular phase diagram (Gibbs's triangle).
    The parameters are same as those in Fig.~\ref{fig:phase_diagram_small_chi}.
    }
    \label{fig:phase_diagram_small_chi_triangle}
 \end{center}
\end{figure}
\begin{figure}[htb]
  \begin{center}
    \includegraphics[width=7cm]{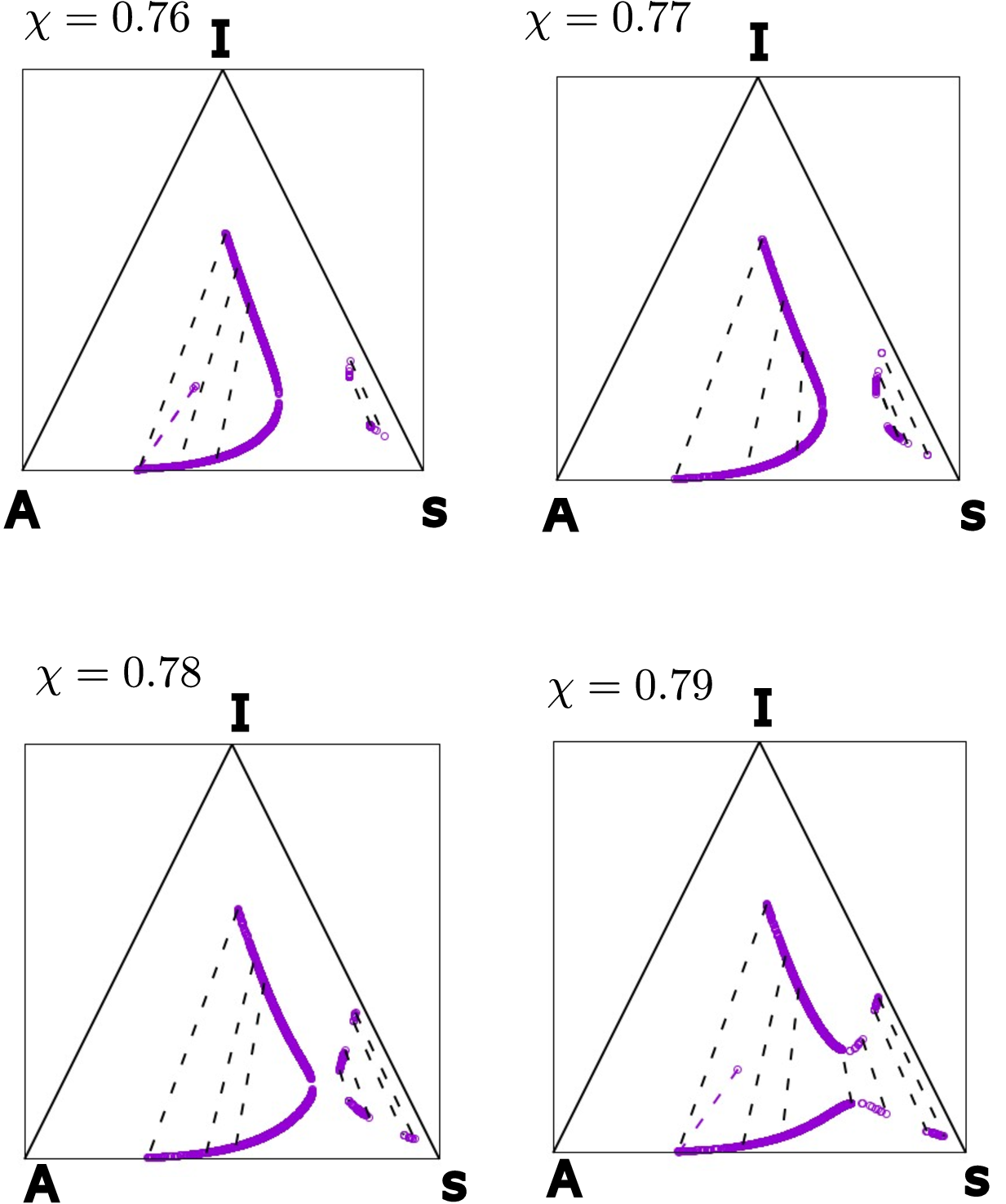}
    \caption{Binodal line described on the triangular phase diagram (Gibbs's triangle).
    The parameters are same as those in Fig.~\ref{fig:phase_diagram_large_chi}.
    }
    \label{fig:phase_diagram_large_chi_triangle}
 \end{center}
\end{figure}
%spinodal_FHMS_triangle
%
\begin{figure}[htb]
  \begin{center}
    \includegraphics[width=6cm]{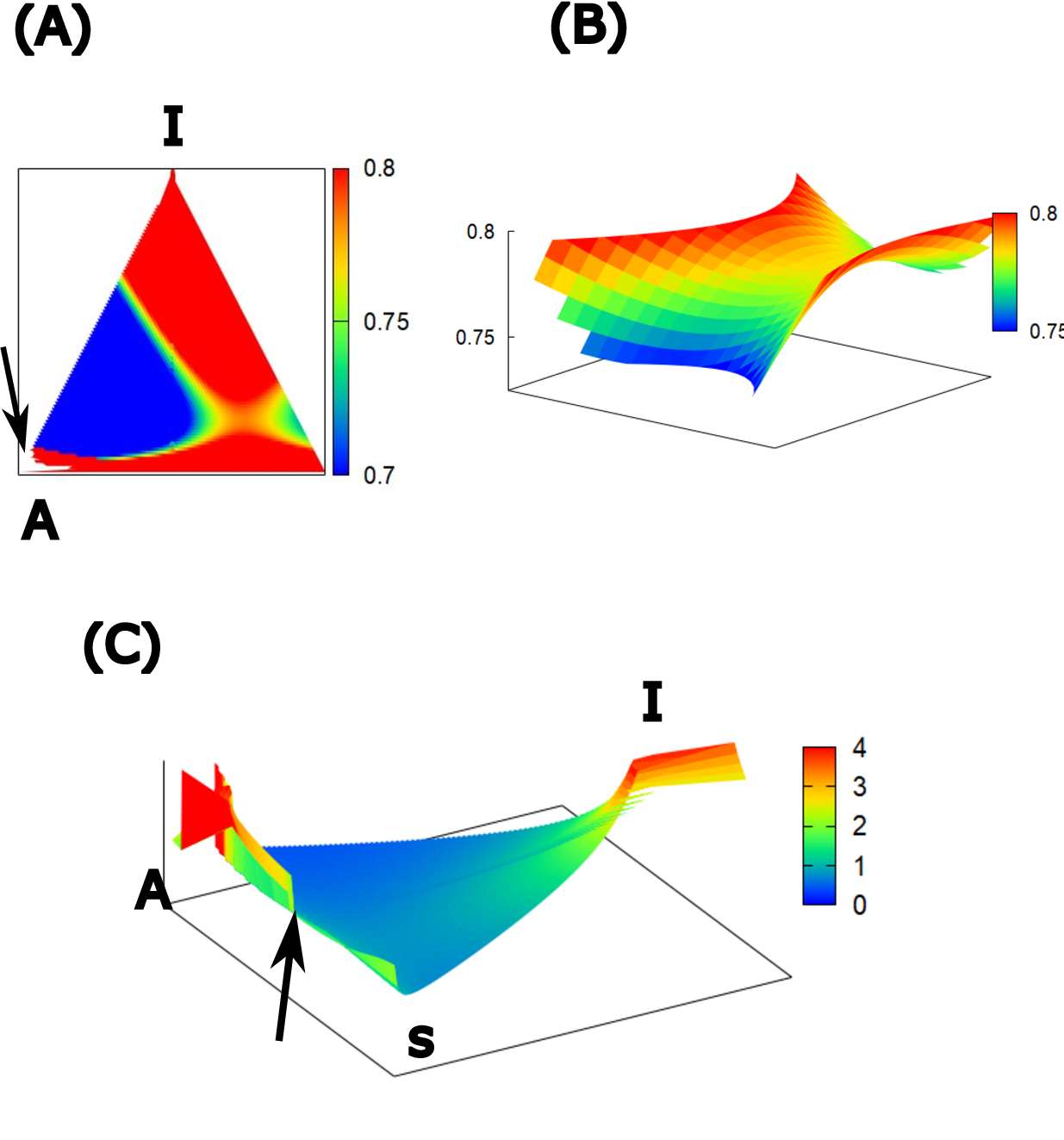}
    \caption{Spinodal surface described on the triangular phase diagram (Gibbs's triangle).
    The parameters are same as those in Fig.~\ref{fig:spinodal}.
    }
    \label{fig:spinodal_triangle}
 \end{center}
\end{figure}
%spinodal_FHMS_triangle

%
The triangular phase diagram (Gibbs triangle) is also a useful method for intuitively visualizing the state specified by the values of $\phi_{\rm I}$ and $\phi_{\rm A}$.
The three vertices of the triangle represent single-component states (only I-, A-, or s-component), while the edges correspond to binary states that exclude the component represented by the opposite vertex.
For example, the bottom edge in Fig.~\ref{fig:phase_diagram_small_chi_triangle} corresponds to the binary mixture composed of A-component and s-component. 
Such triangular phase diagrams are widely used for ternary systems composed of water, oil, and surfactant~\cite{CL}.

In the phase diagrams shown in Figs.~\ref{fig:phase_diagram_small_chi_triangle} and ~\ref{fig:phase_diagram_large_chi_triangle}, it is clear that the binodal line between I-rich and s-rich phases appears after that between I-rich and A-rich phases. 
Moreover, in Fig.~\ref{fig:spinodal_triangle}, the discontinuous region clearly appears in the state specified by a large amount of A-component.
\section{Typical spinodal decomposition pattern for Flory-Huggins theory for a ternary system}\label{sec:typical_spinodal}
\begin{figure}[htb]
  \begin{center}
    \includegraphics[width=9cm]{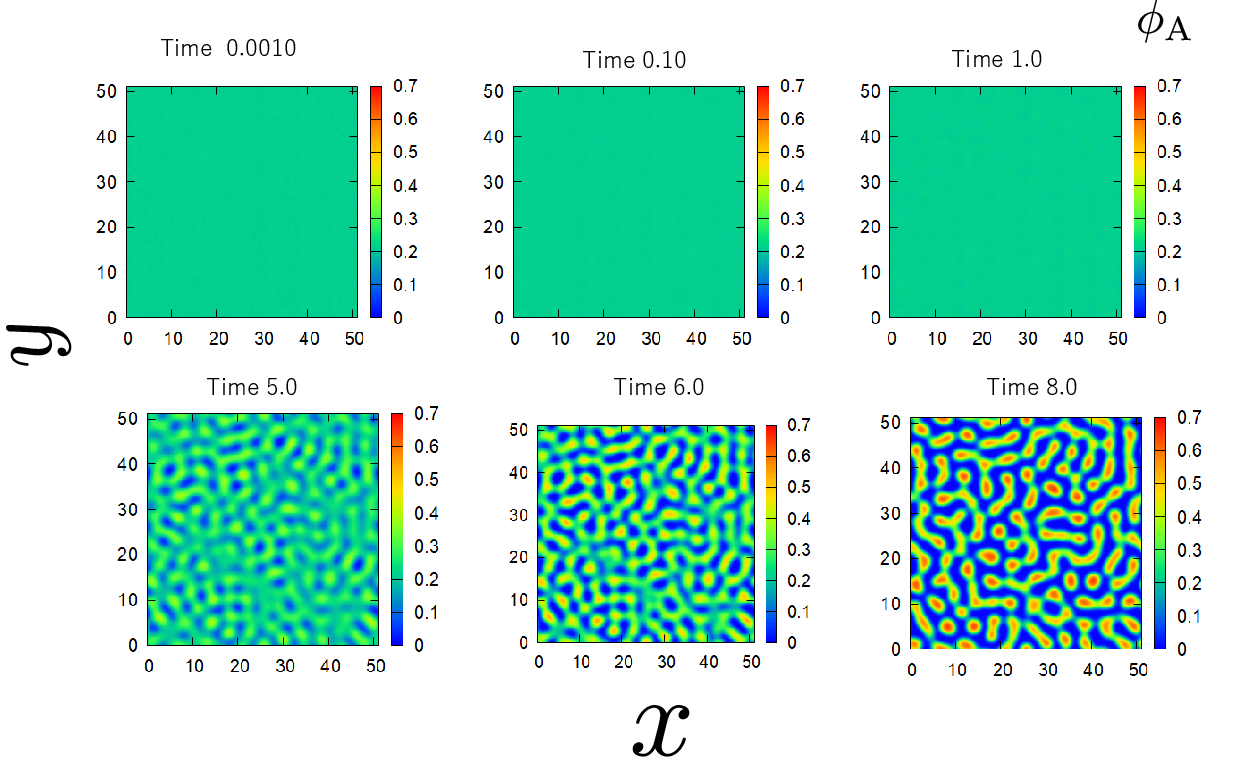}
    \caption{Time evolution of the A-component density field for a typical spinodal decomposition.
    }
    \label{fig:typical_spinodal}
 \end{center}
\end{figure}
For comparison with Fig.~\ref{fig:droplet_formation}, Fig.~\ref{fig:typical_spinodal} shows the typical spinodal decomposition pattern and its time evolution.
The droplet formation shown in Fig.~\ref{fig:droplet_formation} proceeds remarkably faster than the spinodal decomposition shown in Fig.~\ref{fig:typical_spinodal}.
The parameters are $\phi_{\rm I}(t=0)=0.21$, $\phi_{\rm A}(t=0)=0.21$ and $\chi=1.10$.
Note that in the ternary Flory-Huggins theory, the expected growth of the density fluctuation does not appear before time = 5.00  for a parameter set $(\phi_{\rm I}(t=0), \phi_{\rm A}(t=0), \chi)=(0.10, 0.70. 0.75)$ which is the same as Fig.~\ref{fig:droplet_formation}

\section{Power spectrum of density fluctuation during rapid droplet formation}\label{sec:power_spectra}
\begin{figure}[htb]
  \begin{center}
    \includegraphics[width=9cm]{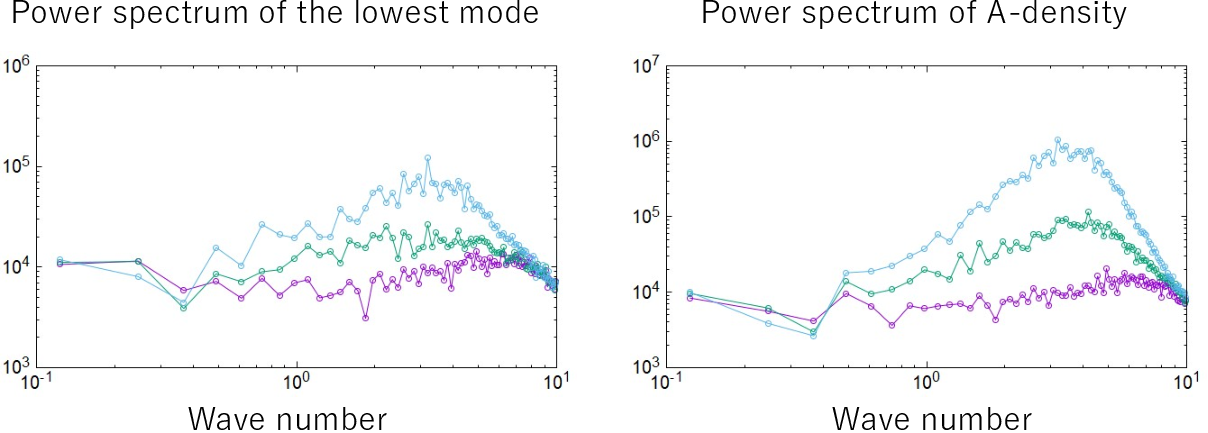}
    \caption{(left) Power spectrum of the smallest eigenmode of Hessian.
    (right)Power spectrum of the density of the A-component.
    Purple, green and cyan indicate the values at $t=0.001, 0.020$ and $0.050$, respectively.
    }
    \label{fig:power_spectra}
 \end{center}
\end{figure}
In order to characteristic the rapid droplet formation, we show the power spectrum of the unstable density fluctuation mode, which is evaluated by the smallest eigenmode of the Hessian matrix defined in eqn.~(\ref{eqn:Hessian}).
As in the case of the spinodal decomposition, the fluctuation of a certain wave number grows with time as shown in Fig.~\ref{fig:power_spectra}.

The power spectra of the smallest eigenmode of the Hessian and the density of the A-component show the spinodal like behavior.
Note that such a spinodal-like behavior is not observed in the simulation without thermal fluctuations.

%\nocite{*}
\bibliographystyle{apsrev4-2}
\bibliography{ref}% Produces the bibliography via BibTeX.

\end{document}